\documentclass[a4paper,12pt]{article}
\usepackage{amsmath,amssymb}

\begin{document}
\setcounter{page}{0}

\begin{titlepage}

\bigskip

\begin{center}
{\LARGE\bf
Supersymmetric $CP^N$ Sigma Model\\
\vspace{3mm}on Noncommutative Superspace}

\vspace{10mm}

\bigskip
{\renewcommand{\thefootnote}{\fnsymbol{footnote}}
\large\bf 
Takeo Inami 
and Hiroaki Nakajima\footnote{
E-mail: nakajima@phys.chuo-u.ac.jp}\\
}

\setcounter{footnote}{0}
\bigskip

{\small \textit{
Department of Physics, Chuo University\\
Kasuga, Bunkyo-ku, Tokyo 112-8551, Japan.}\\
}

\end{center}
\bigskip


\begin{abstract}
We construct a closed form of the action of the supersymmetric $CP^N$ sigma model on noncommutative superspace 
in four dimensions.
We show that this model has $\mathcal{N}=\frac{1}{2}$ supersymmetry and that the transformation law is not modified. 
The supersymmetric $CP^N$ sigma model on noncommutative superspace in two dimensions is obtained by dimensionally reducing 
the model in four dimensions.
\end{abstract}

\end{titlepage}


\section{Introduction}

Noncommutative geometry~\cite{Connes:1997cr} appears in M-theory, string theory and condensed matter physics. 
Noncommutative field theories are known to describe the effective theory of string in a constant NS-NS $B$ field~\cite{Seiberg:1999vs}. 
(2+1)-dimensional noncommutative field theories have been applied to the quantum Hall effect.

In supersymmetric field theories, 
there are a few alternatives in introducing non(anti)commutativity
of the supercoordinates 
$(x^{\mu},\theta^{\alpha} ,\bar{\theta}^{\dot{\alpha}})$\footnote{~We follow the notation of~\cite{Wess:cp}. }. In particular, 
supersymmetric Yang-Mills theory on \textit{noncommutative superspace}
\cite{Klemm:2001yu, deBoer:2003dn, Seiberg:2003yz} describes the effective field theory of string in 
a constant selfdual graviphoton background
\cite{Seiberg:2003yz, Ooguri:2003qp, Berkovits:2003kj}. 
In the field theoritical view point, these theories keep $\mathcal{N}=\frac{1}{2}$ supersymmetry and 
have some interesting properties.

In this letter, we construct the supersymmetric nonlinear sigma model whose target space is $CP^N$ ($CP^N$ SNLSM)
on noncommutative superspace in four and two dimensions. Low-dimensional SNLSMs on ordinary superspace have interesting properties. 
In two dimensions, the $CP^N$ SNLSM is integrable, \textit{i.e.}, it has infinitely many conservation laws. 
It shares important properties with four-dimensional supersymmetric gauge theories, 
such as asymptotic freedom and dynamical mass gap.
In three dimensions, the $CP^N$ SNLSM has been investigated using the large-$N$ expansion~\cite{Inami:2000eb}. 
As we will see below, since the K\"ahler potential of SNLSM is generally non-polynomial, 
the action of SNLSM on noncommutative superspace has infinitely many terms~\cite{Chandrasekhar:2003uq}. 
It is difficult to study the properties of this model either perturbatively or non-perturbatively. 
We introduce an auxiliary vector superfield to linearize the $CP^N$ SNLSM, mimicking the commutative case~\cite{Cremmer:1978bh}.
Once introducing the vector superfield, we can eliminate all auxiliary fields and obtain a closed form of the action.


\section{Noncommutative Superspace}

\subsection{Noncommutative Superspace}

We recapitulate noncommutative supersupace, closely 
following Seiberg~\cite{Seiberg:2003yz}. 
We consider four-dimensional $\mathcal{N}=1$ supersymmetric field theories on the noncommutative superspace. 
The non(anti)commutativity is introduced by 
\begin{gather}
\{\theta^{\alpha} , \theta^{\beta}\}=C^{\alpha\beta},
\quad \{\theta^{\alpha} , \bar{\theta}^{\dot{\alpha}}\}=\{\bar{\theta}^{\dot{\alpha}} , \bar{\theta}^{\dot{\beta}}\}=0,\notag\\
[y^\mu , y^\nu]=[y^\mu , \theta^{\alpha}]=[y^\mu , \bar{\theta}^{\dot{\alpha}}]=0,\label{NC}
\end{gather}
where $y^\mu$ is the chiral coordinate 
\begin{equation}
y^\mu =x^\mu +i\theta\sigma^\mu \bar{\theta}.
\end{equation}
The product of functions of $\theta$ is Weyl ordered by using the Moyal product, which is defined by 
\begin{equation}
\begin{split}
f(\theta)\ast g(\theta)&=f(\theta)\exp\left(-\frac{1}{2}C^{\alpha\beta}\overleftarrow{\frac{\partial}{\partial \theta^{\alpha}}}
                          \overrightarrow{\frac{\partial}{\partial \theta^{\beta}}}\right) g(\theta)\\
                       &=f(\theta)\left[1-\frac{1}{2}C^{\alpha\beta}\overleftarrow{\frac{\partial}{\partial \theta^{\alpha}}}
                          \overrightarrow{\frac{\partial}{\partial \theta^{\beta}}}
                          -\det C \overleftarrow{\frac{\partial}{\partial (\theta\theta)}}
                          \overrightarrow{\frac{\partial}{\partial (\theta\theta)}}\right] g(\theta).
\end{split}
\end{equation}

The supercovariant derivatives are defined by 
\begin{gather}
D_{\alpha}=\frac{\partial}{\partial \theta^{\alpha}}+2i\sigma^{\mu}_{\alpha\dot{\alpha}}\bar{\theta}^{\dot{\alpha}}
\frac{\partial}{\partial y^{\mu}},
\qquad \bar{D}_{\dot{\alpha}}=-\frac{\partial}{\partial \bar{\theta}^{\dot{\alpha}}}.\\
\intertext{Since $D_{\alpha}$ and $\bar{D}_{\dot{\alpha}}$ do not contain $\theta$, their anticommutation relations 
are same as those on the commutative superspace.}
\{D_{\alpha},D_{\beta}\}=0,\quad \{\bar{D}_{\dot{\alpha}},\bar{D}_{\dot{\beta}}\}=0,\quad 
\{D_{\alpha},\bar{D}_{\dot{\alpha}}\}=-2i\sigma^{\mu}_{\alpha\dot{\alpha}}\frac{\partial}{\partial y^{\mu}}.
\end{gather}
The supercharges are defined by 
\begin{gather} 
Q_{\alpha}=\frac{\partial}{\partial \theta^{\alpha}},
\qquad \bar{Q}_{\dot{\alpha}}=-\frac{\partial}{\partial \bar{\theta}^{\dot{\alpha}}}
+2i\theta^{\alpha}\sigma^{\mu}_{\alpha\dot{\alpha}}\frac{\partial}{\partial y^{\mu}}.\\
\intertext{Since $\bar{Q}_{\dot{\alpha}}$ contains $\theta$, the anticommutation relations are modified as follows.}
\{Q_{\alpha},Q_{\beta}\}=0,\qquad \{Q_{\alpha},\bar{Q}_{\dot{\alpha}}\}=+2i\sigma^{\mu}_{\alpha\dot{\alpha}}
\frac{\partial}{\partial y^{\mu}},\\
\{\bar{Q}_{\dot{\alpha}},\bar{Q}_{\dot{\beta}}\}=-4C^{\alpha\beta}\sigma^{\mu}_{\alpha\dot{\alpha}}\sigma^{\nu}_{\beta\dot{\beta}}
\frac{\partial^2}{\partial y^{\mu} \partial y^{\nu}}.\label{q}
\end{gather}
Furthermore, $\bar{Q}_{\dot{\alpha}}$ does not act as derivations on the Moyal product of fields 
\begin{equation}
\bar{Q}_{\dot{\alpha}}(f\ast g)\neq (\bar{Q}_{\dot{\alpha}}f)\ast g +f\ast(\bar{Q}_{\dot{\alpha}}g).
\end{equation}
Then $\bar{Q}_{\dot{\alpha}}$ is not a symmetry of the theory in general, hence we have $\mathcal{N}=\frac{1}{2}$ supersymmetry.

\subsection{Superfields}

The chiral superfield is defined by $\bar{D}_{\dot{\alpha}}\Phi=0$, and hence, $\Phi=\Phi(y,\theta)$.
In terms of the component fields, it is given by 
\begin{equation}
\Phi(y,\theta)=\phi(y)+\sqrt{2}\theta\psi(y)+\theta\theta F(y),
\end{equation}
where $\theta\theta=-\theta^{1}\theta^{2}+\theta^{2}\theta^{1}$ and is Weyl ordered. 

The antichiral superfield is defined by $D_{\alpha}\bar{\Phi}=0$, 
and hence, $\bar{\Phi}=\bar{\Phi}(\bar{y},\bar{\theta})$, where $\bar{y}^{\mu}$ is given by 
\begin{gather}
\bar{y}^{\mu}=y^{\mu}-2i\theta\sigma^\mu \bar{\theta},\\
\quad [\bar{y}^{\mu},\bar{y}^{\nu}]=4\bar{\theta}\bar{\theta}C^{\mu\nu},
\quad C^{\mu\nu}=C^{\alpha\beta}\epsilon_{\beta\gamma}(\sigma^{\mu\nu})_{\alpha}^{~\,\gamma}.
\end{gather}
In the component fields, it is convenient to express the antichiral superfield in terms of $y$ and $\theta$ 
and to Weyl order the $\theta$s 
\begin{equation}
\begin{split}
\bar{\Phi}(y-2i\theta\sigma\bar{\theta},\bar{\theta})
          &=\bar{\phi}(y)+\sqrt{2}\bar{\theta}\bar{\psi}(y)-2i\theta\sigma^\mu \bar{\theta}\partial_{\mu}\bar{\phi}(y) \\
          &\qquad +\bar{\theta}\bar{\theta}\Bigl[\bar{F}(y)+\sqrt{2}i\theta\sigma^\mu \partial_{\mu}\bar{\psi}(y)
           +\theta\theta\partial^{2}\bar{\phi}(y)\Bigr].
\end{split}
\end{equation}

We also need the U(1) vector superfield in constructing the $CP^N$ SNLSM later. 
The vector superfield is written in the Wess-Zumino gauge as 
\begin{align}
V(y,\theta,\bar{\theta})&=-\theta\sigma^\mu \bar{\theta}A_{\mu}(y)+i\theta\theta\bar{\theta}\bar{\lambda}(y)
                         -i\bar{\theta}\bar{\theta}\theta^{\alpha}\Biggl[\lambda_{\alpha}(y)+
                         \frac{1}{4}\epsilon_{\alpha\beta}C^{\beta\gamma}\sigma^{\mu}_{\gamma\dot{\gamma}}
                         \{\bar{\lambda}^{\dot{\gamma}},A_{\mu}\}\Biggr]\notag\\
                        &\qquad +\frac{1}{2}\theta\theta\bar{\theta}\bar{\theta}\bigl[D(y)-i\partial_{\mu}A^{\mu}(y)\bigr].
\end{align}
The $C$-deformed part in the $\bar{\theta}\bar{\theta}\theta$ term is introduced in order that the component fields transform 
canonically under the gauge transformation. The powers of $V$ are obtained by 
\begin{align}
V^{2}&=\bar{\theta}\bar{\theta}\Biggl[-\frac{1}{2}\theta\theta A_{\mu}A^{\mu}-\frac{1}{2}C^{\mu\nu}A_{\mu}A_{\nu}\notag\\
     & \qquad\qquad +\frac{i}{2}\theta_{\alpha}C^{\alpha\beta}\sigma^{\mu}_{\beta\dot{\alpha}}[A_{\mu},\bar{\lambda}^{\dot{\alpha}}]
     -\frac{1}{8}|C|^{2}\bar{\lambda}\bar{\lambda}\Biggr],\\
V^{3}&=0,
\end{align}
where $|C|^{2}=C^{\mu\nu}C_{\mu\nu}=4\det C$.

\section{Supersymmetric $CP^N$ Sigma Model on \\Noncommutative Superspace}


The Lagrangian of four-dimensional $\mathcal{N}=1$ 
supersymmetric nonlinear sigma model (SNLSM) is 
written using the K\"ahler potential $K$ as 
\begin{equation}
\mathcal{L}=\int \! d^{2}\theta d^{2}\bar{\theta}\,K(\Phi,\bar{\Phi}).\label{K}
\end{equation}
The same expression can be used for $\mathcal{N}=2$ SNLSM in two dimensions.
The results derived below for the four-dimensional case hold true for the two-dimensional case 
after slight modification. See the argument of dimensional reduction to two dimensions in the end of this section.

In the case of a single pair of chiral and antichiral superfields, the Berezin integration in 
eq.\eqref{K} with the noncommutativity \eqref{NC} was calculated in~\cite{Chandrasekhar:2003uq}. It is given by 
\begin{equation}
\mathcal{L}=\mathcal{L}(C=0)+\sum_{n=1}^{\infty}(\det C)^{n}[A_{n}F^{2n-1}+B_{n}F^{2n}],\label{K2}
\end{equation}
where $A_{n}$ and $B_{n}$ are functions of the component fields. Eq.\eqref{K2} contains infinitely many terms 
since generally $K$ is not a polynomial and the powers of $\theta$ are nonzero. 
We have not found a good way to analyze this model as it stands.

Some SNLSMs are expressed as supersymmetric gauge theories. Such SNLSMs contain the model whose target space is a 
Hermitian symmetric space~\cite{Higashijima:1999ki} (\textit{e.g.} $CP^N$ and Grassmannian 
$G_{N,M}$), $T^{\ast}CP^N$~\footnote{~$T^{\ast}CP^N$ denotes the cotangent bundle of $CP^N$. $T^{\ast}G_{N,M}$ is similar. } 
and $T^{\ast}G_{N,M}$. 
We construct the $CP^N$ SNLSM on noncommutative superspace as the noncommutative extension of~\cite{Cremmer:1978bh} 
using the result of~\cite{Araki:2003se}. 
We start from the following Lagrangian
\begin{equation}
\mathcal{L}=\int \! d^{2}\theta d^{2}\bar{\theta}\,\Bigl[\bar{\Phi}^{i} \ast e^V \ast \Phi^{i} - V\Bigr],
\end{equation}
where $i=1,2,\ldots ,N+1$. $V$ is the U(1) vector superfield. It is written modulo total derivatives
in terms of component fields as 
\begin{equation}
\begin{split}
\mathcal{L}&=\bar{F}^{i} F^{i} -i\bar{\psi}^{i} \bar{\sigma}^{\mu} \mathcal{D}_\mu \psi^{i} 
             -\mathcal{D}_\mu \bar{\phi}^{i} \mathcal{D}^{\mu} \phi^{i} +\frac{1}{2}\bar{\phi}^{i} D \phi^{i} 
             +\frac{i}{\sqrt{2}}(\bar{\phi}^{i} \lambda\psi^{i}-\bar{\psi}^{i} \bar{\lambda}\phi^{i})\cr
           & \quad~ +\frac{i}{2}C^{\mu\nu}\bar{\phi}^{i}F_{\mu\nu}F^{i}-\frac{1}{16}|C|^2 \bar{\phi}^{i}\bar{\lambda}\bar{\lambda}F^{i} 
                   -\frac{1}{\sqrt{2}}C^{\alpha\beta}(\mathcal{D}_{\mu} \bar{\phi}^{i})\sigma^{\mu}_{\beta\dot{\alpha}} 
                   \bar{\lambda}^{\dot{\alpha}} \psi^{i}_{\alpha}\cr
           & \quad~ -\frac{1}{2}D.
\end{split}\label{Laux}
\end{equation}
Here $\mathcal{D}_{\mu}$ is the gauge covariant derivative defined by 
\begin{equation}
\mathcal{D}_\mu \phi^{i}=\partial_{\mu}\phi^{i}+\frac{i}{2}A_{\mu}\phi^{i},\quad 
\mathcal{D}_\mu \psi^{i}=\partial_{\mu}\psi^{i}+\frac{i}{2}A_{\mu}\psi^{i}.
\end{equation}
Following~\cite{Araki:2003se}, we redefine the antichiral superfields $\bar{\Phi}^{i}$ in the Lagrangian \eqref{Laux} as
\begin{equation}
\begin{split}
\bar{\Phi}^{i}(\bar{y},\bar{\theta})&=\bar{\phi}^{i}(\bar{y})+\sqrt{2}\bar{\theta}\bar{\psi}^{i}(\bar{y})\\
&\qquad +\bar{\theta}\bar{\theta}\Bigl[\bar{F}^{i}(\bar{y})+iC^{\mu\nu}\partial_{\mu}(\bar{\phi}^{i}A_{\nu})(\bar{y})
-\frac{1}{4}C^{\mu\nu}\bar{\phi}^{i}A_{\mu}A_{\nu}(\bar{y})\Bigr],
\end{split}
\end{equation}
so that the component fields transform canonically under the gauge transformation. 

Eq.\eqref{Laux} contains the auxiliary fields $F^{i}$, $\bar{F}^{i}$, $D$, $\lambda$, $\bar{\lambda}$ and $A_{\mu}$. 
They have the role of imposing constraints on the fields as follows. 
\begin{eqnarray}
\bar{F}^{i} &:& F^{i}=0, \\
F^{i}&:&\bar{F}^{i}+\frac{i}{2}C^{\mu\nu}\bar{\phi}^{i}F_{\mu\nu}
-\frac{1}{16}|C|^{2}\bar{\phi}^{i}\bar{\lambda}\bar{\lambda}=0,\\
D~ &:& \bar{\phi}^{i} \phi^{i}=1, \\
\lambda^{\alpha}&:&\bar{\phi}^{i}\psi^{i}_{\alpha}=0,\\
\bar{\lambda}^{\dot{\alpha}}&:&\frac{i}{\sqrt{2}}\bar{\psi}^{i}_{\dot{\alpha}}\phi^{i}
-\frac{1}{8}|C|^2 \bar{\phi}^{i} \bar{\lambda}_{\dot{\alpha}}F^{i}-\frac{1}{\sqrt{2}}C^{\alpha\beta}(\mathcal{D}_{\mu} \bar{\phi}^{i})
\sigma^{\mu}_{\beta\dot{\alpha}} \psi^{i}_{\alpha}=0, \\
A_{\mu}&:&\frac{1}{2}\bar{\psi}^{i} \bar{\sigma}^{\mu} \psi^{i}
+\frac{i}{2}(\bar{\phi}^{i}\partial^{\mu} \phi^{i}-\partial^{\mu} \bar{\phi}^{i} \cdot \phi^{i})
-\frac{1}{2}(\bar{\phi}^{i} \phi^{i})A^{\mu}\nonumber \\
&& \qquad +iC^{\mu\nu}\partial_{\nu}(\bar{\phi}^{i}F^{i})-\frac{i}{2\sqrt{2}}C^{\alpha\beta}\sigma^{\mu}_{\beta\dot{\alpha}} 
\bar{\lambda}^{\dot{\alpha}} \bar{\phi}^{i} \psi^{i}_{\alpha}=0.
\end{eqnarray}
After eliminating $F^{i}$ and $\bar{F}^{i}$, the Lagrangian \eqref{Laux} takes a simple form 
\begin{gather}
\mathcal{L}=-\mathcal{D}_\mu \bar{\phi}^{i} \mathcal{D}^{\mu} \phi^{i} -i\bar{\psi}^{i} \bar{\sigma}^{\mu} \mathcal{D}_\mu \psi^{i},
\label{L2}\\
\intertext{with the constraints}
\bar{\phi}^{i} \phi^{i}=1,\label{c1}\\
\bar{\phi}^{i}\psi^{i}_{\alpha}=0,\label{c2}\\
\bar{\psi}^{i}_{\dot{\alpha}}\phi^{i}+iC^{\alpha\beta}\sigma^{\mu}_{\beta\dot{\alpha}}(\mathcal{D}_{\mu} \bar{\phi}^{i})
\psi^{i}_{\alpha}=0,\label{c3}\\
A_{\mu}=i(\bar{\phi}^{i}\partial_{\mu} \phi^{i}-\partial_{\mu} \bar{\phi}^{i} \cdot \phi^{i})
+\bar{\psi}^{i}\bar{\sigma}^{\mu}\psi^{i}.\label{c4}
\end{gather}

The constraints (\ref{c1}-\ref{c3}) are solved as follows 
\begin{align}
\phi^{i}&=\frac{1}{\sqrt{1+\bar{\varphi}\varphi}}\binom{\varphi^{a}}{1}, & 
\bar{\phi}^{i}&=\frac{1}{\sqrt{1+\bar{\varphi}\varphi}}\binom{\bar{\varphi}^{\bar{a}}}{1},\label{phi}\\[3mm]
\psi^{i}_{\alpha}&=\frac{1}{\sqrt{1+\bar{\varphi}\varphi}}P^{ij}\chi^{j}_{\alpha}, & 
\chi^{i}_{\alpha}&=\binom{\chi^{a}_{\alpha}}{0},\label{chi}
\end{align}
\begin{equation}
\bar{\psi}^{i}_{\dot{\alpha}}=\frac{1}{\sqrt{1+\bar{\varphi}\varphi}}\Bigl[\bar{\chi}^{j}_{\dot{\alpha}}P^{ji}
-iC^{\alpha\beta}\sigma^{\mu}_{\beta\dot{\alpha}}\bar{\phi}^{i}(\partial_{\mu}\bar{\phi}^{j})P^{jk} \chi^{k}_{\alpha}\Bigr],
\quad \bar{\chi}^{i}_{\dot{\alpha}}=\binom{\bar{\chi}^{\bar{a}}_{\dot{\alpha}}}{0},\label{chibar}
\end{equation}
where $a,\bar{a}=1,2,\ldots,N$. $P^{ij}=\delta^{ij}-\phi^{i}\bar{\phi}^{j}$ is a projection operator which satisfies 
\begin{equation}
P^2 =P,\quad \bar{\phi}^{i}P^{ij}=P^{ij}\phi^{j}=0.
\end{equation}
Substituting eqs.(\ref{c4}-\ref{chibar}) into eq.\eqref{L2}, the Lagrangian becomes 
\begin{align}
\mathcal{L}&=\mathcal{L}_{0}+\mathcal{L}_{C}, \label{LL}\\
\mathcal{L}_{0}&=-g_{a\bar{b}}\partial_{\mu}\varphi^{a}\partial^{\mu}\bar{\varphi}^{\bar{b}}
             -ig_{a\bar{b}}\bar{\chi}^{\bar{b}}\bar{\sigma}^{\mu}D_{\mu}\chi^{a}
             -\frac{1}{4}R_{a\bar{b}c\bar{d}}(\chi^{a}\chi^{c})(\bar{\chi}^{\bar{b}}\bar{\chi}^{\bar{d}}), \\
\mathcal{L}_{C}&=2g_{a\bar{b}}g_{c\bar{d}}\,C^{\alpha\beta}(\sigma^{\mu\nu})_{\beta}^{~\,\gamma}\chi^{a}_{\alpha}\chi^{c}_{\gamma}
             (\partial_{\mu}\bar{\varphi}^{\bar{b}})(\partial_{\nu}\bar{\varphi}^{\bar{d}}),
\end{align}
where $g_{a\bar{b}}$, $D_{\mu}\chi^{a}$ and $R_{a\bar{b}c\bar{d}}$ are given by 
\begin{gather}
g_{a\bar{b}}=\frac{(1+\bar{\varphi}\varphi)\delta_{ab}-\bar{\varphi}^{\bar{a}}\varphi^{b}}{(1+\bar{\varphi}\varphi)^2},\quad 
D_{\mu}\chi^{a}=\partial_{\mu}\chi^{a}+\varGamma^{a}_{bc}(\partial_{\mu}\varphi^{b})\chi^{c},\\
\varGamma^{a}_{bc}\equiv g^{a\bar{d}}\partial_{b}g_{c\bar{d}}, \quad
R_{a\bar{b}c\bar{d}}\equiv -g_{a\bar{e}}\partial_{c}(g^{f\bar{e}}\partial_{\bar{d}}g_{f\bar{b}})
=g_{a\bar{b}}g_{c\bar{d}}+g_{a\bar{d}}g_{c\bar{b}}.
\end{gather}
$g_{a\bar{b}}$ is the Fubini-Study metric of $CP^N$. $\varGamma^{a}_{bc}$ and $R_{a\bar{b}c\bar{d}}$ are the Christoffel symbol and 
the Riemann curvature tensor respectively. In the $CP^1$ case, the $C$-deformed part $\mathcal{L}_{C}$ vanishes.
\begin{equation}
\mathcal{L}_{C}^{(CP^{1})}=2(1+\bar{\varphi}\varphi)^{-4}C^{\alpha\beta}(\sigma^{\mu\nu})_{\beta}^{~\,\gamma}
                             \chi_{\alpha}\chi_{\gamma}(\partial_{\mu}\bar{\varphi})(\partial_{\nu}\bar{\varphi})=0.
\end{equation}

We study supersymmetry of the Lagrangian \eqref{LL}. In the $C=0$ case, the $\mathcal{N}=1$ supersymmetry transformation 
is generated by $Q_{\alpha}$ and $\bar{Q}_{\dot{\alpha}}$. $Q_{\alpha}$ generates the transformation 
\begin{align}
\delta \varphi^{a}&=\sqrt{2}\,\xi\chi^{a},& \delta \bar{\varphi}^{\bar{a}}&=0,\label{tr1}\\
\delta \chi^{a}_{\alpha}&=-\sqrt{2}\varGamma^{a}_{bc}(\xi\chi^{b})\chi^{c}_{\alpha},& 
\delta \bar{\chi}^{\bar{a}}_{\dot{\alpha}}&=-\sqrt{2}i(\bar{\sigma}^{\mu}\xi)_{\dot{\alpha}}\partial_{\mu}\bar{\varphi}^{\bar{a}}.\label{tr2}
\end{align}
In the present case with $C\neq 0$, the same transformation \eqref{tr1} and \eqref{tr2} give
\begin{align}
\delta g_{a\bar{b}}&=\partial_{c}g_{a\bar{b}}\,\delta\varphi^{c}+\partial_{\bar{c}}g_{a\bar{b}}\,\delta\varphi^{\bar{c}}=
\sqrt{2}g_{b\bar{b}}\varGamma^{b}_{ac}(\xi\chi^{c}),\\
\delta(g_{a\bar{b}}\chi^{a}_{\alpha}\partial_{\mu}\bar{\varphi}^{\bar{b}})&=\Bigl[\sqrt{2}g_{b\bar{b}}\varGamma^{b}_{ac}(\xi\chi^{c})\chi^{a}_{\alpha}
-\sqrt{2}\varGamma^{a}_{bc}(\xi\chi^{b})\chi^{c}_{\alpha}g_{a\bar{b}}\Bigr]\partial_{\mu}\bar{\varphi}^{\bar{b}}=0.\\
\intertext{We then obtain}
\delta \mathcal{L}_{C}&=\delta\Bigl[2C^{\alpha\beta}(\sigma^{\mu\nu})_{\beta}^{~\,\gamma}
(g_{a\bar{b}}\chi^{a}_{\alpha}\partial_{\mu}\bar{\varphi}^{\bar{b}})(g_{c\bar{d}}\chi^{c}_{\gamma}\partial_{\nu}\bar{\varphi}^{\bar{d}})
\Bigr]=0.
\end{align}
We have shown that the Lagrangian \eqref{LL} is invariant under the $\mathcal{N}=\frac{1}{2}$ supersymmetry transformation 
\eqref{tr1} and \eqref{tr2}. 

Using dimensional reduction, we obtain the Lagrangian of $CP^N$ SNLSM on noncommutative superspace in two dimensions.
\begin{align}
\mathcal{L}_{2D}&=\frac{1}{2}g_{AB}\partial_{z}\varphi^{A}\partial_{\bar{z}}\varphi^{B}
             +ig_{a\bar{b}}\Bigl(\bar{\chi}^{\bar{b}}_{L}D_{\bar{z}}\chi^{a}_{L}+\bar{\chi}^{\bar{b}}_{R}D_{z}\chi^{a}_{R}\Bigr)
             +R_{a\bar{b}c\bar{d}}\chi^{a}_{L}\bar{\chi}^{\bar{b}}_{L}\chi^{c}_{R}\bar{\chi}^{\bar{d}}_{R} \notag\\
           & \qquad\qquad +2g_{a\bar{b}}g_{c\bar{d}}\,(C^{11}\chi^{a}_{L}\chi^{c}_{L}-C^{22}\chi^{a}_{R}\chi^{c}_{R})
             \epsilon^{\mu\nu}(\partial_{\mu}\bar{\varphi}^{\bar{b}})(\partial_{\nu}\bar{\varphi}^{\bar{d}}),
\end{align}
where
\begin{equation}
\varphi^{A}=(\varphi^{a},~\bar{\varphi}^{\bar{a}}), \quad 
g_{AB}=\begin{pmatrix} 0 & g_{a\bar{b}} \\ g_{b\bar{a}} & 0 \end{pmatrix}, \quad
\chi^{a}_{\alpha}=\begin{pmatrix} \chi^{a}_{L} \\ \chi^{a}_{R} \end{pmatrix}, \quad 
\bar{\chi}^{\bar{a}}_{\dot{\alpha}}=\begin{pmatrix}  \bar{\chi}^{\bar{a}}_{L} \\ \bar{\chi}^{\bar{a}}_{R} \end{pmatrix}.
\end{equation}


\section{Discussion}

In this letter, we have studied the supersymmetric $CP^N$ sigma model on noncommutative superspace. 
We have constructed a closed form of the Lagrangian of the model \eqref{LL}. We have found that the $\mathcal{N}=\frac{1}{2}$ 
supersymmetry transformation law of the model is not modified.

In two dimensions ordinary NLSMs with extended supersymmetry have a few remarkable properties.
\begin{description}
\item[~~i)] Models are integrable (at least at classical level).
\item[~ii)] They have good UV divergence properties, \textit{i.e.}, finite to certain loops for $\mathcal{N}=2$ and 
finite to all loops for $\mathcal{N}=4$.
\item[iii)] They possess instantons. 
\end{description}
It is interesting to see whether these nice properties hold true for two-dimensional SNLSM on noncommutative superspace.

\medskip

\noindent {\bf Acknowledgements}

We thank Chong-Sun Chu, Ko Furuta, Takeo Araki, Katsushi Ito and Muneto Nitta for useful conversations and comments. 
This work is supported partially by Chuo University grant for special research and 
research grants of Japanese ministry of Education and Science, Kiban C and Kiban B.


\end{document}